\documentstyle[twoside,fleqn,espcrc2,epsfig]{article}

\def\beq{\begin{equation}}
\def\eeq{\end{equation}}
\def\bea{\begin{eqnarray}}
\def\eea{\end{eqnarray}}

\newcommand{\permil}{\raisebox{0.5ex}{\tiny $0$}$\!/$\raisebox{-0.3ex}{\tiny $\! 00$}{\normalsize \hspace{1ex}}}

\newcommand{\gessim}{\raisebox{0.5ex}{$>$}\hspace{-1.7ex}\raisebox{-0.5ex}{$\sim$}}
\def\cpc#1#2#3{Computer Phys.\ Comm.\ #1 (19#3) #2}

\def\np#1#2#3{Nucl.\ Phys.\ B#1 (19#3) #2}
\def\pl#1#2#3{Phys.\ Lett.\ B#1 (19#3) #2}
\def\pr#1#2#3{Phys.\ Rev.\ D #1 (19#3) #2}

\def\zp#1#2#3{Zeit.\ Phys.\ C#1 (19#3) #2}

\begin{document}

% add words to TeX's hyphenation exception list
\hyphenation{author another created financial paper re-commend-ed}

\pagestyle{empty}

\twocolumn[
%{\bf 
{\Large Experimental studies of QCD using flavour tagged jets from DELPHI} 
%}
\vspace*{0.5cm} \\
J. Fuster$^a$, S. Cabrera$^a$ and S. Mart\'{\i} i Garc\'{\i}a$^b$
\vspace*{0.5cm} \\
      $^a$~IFIC, centre mixte CSIC--Universitat de Val\`encia,
      Avda. Dr. Moliner 50,\\ E--46100 Burjassot,  Val\`encia, Spain, 
      E-mail: fuster@evalo1.ific.uv.es, cabrera@evalo1.ific.uv.es
\vspace*{0.3cm} \\
      $^b$~Department of Physics, University of Liverpool, P.O. Box
      147,\\ 
      Liverpool L69 3BX, United Kingdom, E-mail: martis@hep.ph.liv.ac.uk 
\vspace*{0.3cm} \\
{\small
Identified $b\overline{b}g$ and $q\overline{q}\gamma$ events from DELPHI are used to 
measure the ratio of the mean charged particle multiplicity distribution between 
gluon and quark jets. The dependence of this ratio with the jet energy is
established using about three million Z$^0$ decays.  Results from all other detectors
are discussed and compared. A nice agreement is found among all them. The
ratio between the normalized total three-jet cross sections of
$b\overline{b}g$ and $q\overline{q}g,\ q \equiv u,d,s$ events is also 
determined. The preliminary value obtained indicates that $b$ quarks
are experimentaly seen to radiate less than light quarks due to their higher 
mass. The suggested experimental error is $\sim$300 MeV for the $b$ mass
determination at the M$_Z$ scale.   
}
\vspace*{0.4cm} 
] 

\section{INTRODUCTION}

In Quantum Chromodynamics  (QCD), quarks ($q$) and  gluons ($g$) are coloured
objects that carry different colour charges.  Quarks have  a single  colour 
index  while gluons are tensor objects carrying two colour  indices. Due to 
this fact, quarks and gluons differ in their  relative coupling  strength to 
emit  additional gluons, and, in consequence,  jets originating  from the  
fragmentation of  energetic quarks and gluons are expected to show differences
in their final particle multiplicities, energies  and angular  distributions. 

The masses of quarks are also fundamental parameters of the QCD 
lagrangian not predicted by the theory. The definition of the quark 
masses is however not unique because quarks are not free particles and 
various scenarios are possible. The perturbative pole mass, M$_q$, and the 
running mass, m$_q$, of the $\overline{MS}$ scheme are among the most 
currently used. At first order in $\alpha_s$
the predicted expression for an observable is not able to resolve
the mass ambiguity and, only when second or higher order terms are 
included, the mass definition becomes known. At orders higher than one 
the renormalization scheme used as the baseline of the calculation has 
to be chosen and this contains the information about the mass 
definition. Earlier calculations of the three-jet cross section in 
$e^+e^-$ including mass terms already exist at O($\alpha_s$) \cite{tor,val}
 and have been used to evaluate 
mass effects for the $b$-quark when testing the universality of 
the strong coupling constant, $\alpha_s$. They could not however be 
used to evaluate the mass of the $b$-quark, m$_b$, because these 
calculations are ambiguous in this parameter.  Recently, expressions 
at O($\alpha_s^2$),  for the multi-jet production rate in $e^+e^-$ 
are  available \cite{german} and, thus, they enable  measuring 
m$_b$ in case the flavour independence of $\alpha_s$ is assumed and 
enough experimental precision is achieved.
 
There are well known existing difficulties to measure 
all the above parameters in quantitative  agreement   with the  
predictions  from  perturbative   QCD, since partons, quarks and gluons, are 
not directly observed  in nature and only the stable particles, produced  
after the  fragmentation  process,  are experimentally detected.  However, the 
massive statistics and improved jet tagging techniques available
at LEP presently allow overcomig these difficulties by  applying restrictive 
selection criteria which lead to quark and gluon jet samples with 
high purities. The selected data  samples are almost background  
free and small  corrections to account for impurities are  needed. A  
smaller model dependence than ever is now  achieved, bringing the 
possibility  to perform quantitative studies of quark and gluon 
fragmentation according to perturbative QCD.

The analyses reported in here include more than 3 million Z$^0$ decays as
collected by DELPHI at center-of-mass energies of $\sqrt{s}\approx$M$_Z$.   
In the first analysis, the ratio between the gluon jet multiplicity  and  
the quark jet  multiplicity, $r=\langle N_g \rangle /\langle N_q \rangle$, is 
presented and discussed in comparison with other detector results. In the 
second study, preliminary values of errors associated to the determination 
of m$_b$ at the M$_Z$ scale are given.
 
\section{EVENT  SELECTION}

%%%%%%%%%%%%%%%%%%%%%%%%%%%%%%%%%%%%%%%%%%%%%%%%%%%%%%%%%%%%%%%%%%%%%%%%%%%%%%%%

Gluon and quark jets were selected using hadronic three-jet events. Jets 
were mainly reconstructed  using the {\sc Durham}  algorithm although the 
{\sc  Jade} algorithm was also used \cite{delphi_qg},  in particular, 
to observe the effects due to different  angular  particle  acceptance of 
the  various  algorithms.

In  the  gluon splitting process  ($g\rightarrow  q\bar{q}$), the heavy quark
production is  strongly suppressed \cite{gluo_qq}.  Gluon jets can thus be extracted  
from  $q\bar{q}g$ events by applying $b$ tagging techniques. The two jets
which satisfy the experimental signatures  of being initiated by  $b$ quarks 
are associated to the quark jets and the remaining one is, $by$ $definition$, 
assigned to be the gluon  jet  without any further requirement. Algorithms 
for tagging  $b$ jets exploit the fact that  the decay products of long lived 
B hadrons have large impact parameters and/or contain inclusive  high momentum 
leptons coming from the  semileptonic decays of the  B hadrons.  Gluon  
purities of 94\% and 85\% are achieved when using these techniques,  
respectively. Obviously, the quark jets belonging to these events cannot be 
used to  represent an unbiased quark sample. Thus   the quark jets whose  
properties  are to be  compared with the gluon  jets must  be  selected 
from other   sources which in  any case should preserve  the same
kinematics. Two  possibilities  have been proposed in  the
current   literature.  One consists  in selecting   symmetric three-jet event
configurations \cite{delphi_qg,opal_qg,aleph_qg} in which  one (Y) or the  two
(Mercedes) quark jets have similar energy to that of the gluon jet. The quark
jet  purities reached are $\sim$52\%  and $\sim$66\%, for  Y and for Mercedes
events, respectively. In a second
solution \cite{delphi_qg,opal_qg,l3_qqgamma} radiative $q\bar{q}\gamma$ events
are selected, allowing  a  sample of quark  jets  with variable energy  to be
collected.  In this latter case,  misidentifications of $\gamma$'s due to the
$\pi^\circ$  background and radiative $\tau^+\tau^-\gamma$ contamination give
rise to quark jet  purities of $\sim$92\%.  This method gives a higher purity
but unfortunately suffers from the lack of statistics.

The $b$-quark purity in the $b\bar{b}g$ sample reached in the DELPHI analyses 
is $\sim$93\% and for the light $uds$-quarks is $\sim$80\%. Table 
\ref{tab:events} summarizes the number of events selected and their 
corresponding energy intervals.
  
{\small
\begin{table}[t]
\begin{center}
\caption{The three-jet event samples and their corresponding energy intervals as
used in the analysis.}
\begin{tabular}{lrc}
\hline
Event type & \# events & Jet energy range \\
\hline 
$q\bar{q}\gamma$ & $\begin{array}{r} 2,237 \\ \end{array}$  
& 7.5~GeV - 42.5~GeV \\
$(uds)\overline{(uds)}g$ & $\begin{array}{r} 552,645 \\ \end{array}$  
& 7.5~GeV - 42.5~GeV \\
$b\bar{b}g$    & $\begin{array}{r} 104,081 \\ \end{array}$  
& 7.5~GeV - 42.5~GeV \\
Y  & $\begin{array}{r} 74,164 \\ \end{array}$ 
& 19.6~GeV - 28.8~GeV \\
Mercedes & $\begin{array}{r} 9,264 \\ \end{array}$ &
27.4~GeV - 33.4~GeV \\
\hline
\end{tabular}
\label{tab:events}
\end{center}
\end{table}
}

\section{MULTIPLICITIES OF QUARK AND GLUON JETS}

Results   on    the   charged    multiplicity     of    quark   and     gluon
jets \cite{opal_qg,aleph_qg}  using     symmetric   Y   configurations     and
reconstructed with {\sc Durham} at 24  GeV gluon jet  energy, give a ratio of
$r \approx 1.23  \pm 0.04 \mbox{(stat.+syst.)}$  which does not depend on the
cut-off parameter ($y_{cut}$) selected  to reconstruct jets \cite{opal_qg}. It
is significantly  higher than one, which  indicates that quark  and gluons in
fact fragment differently,  but it  remains far  from the  asymptotic  lowest
order expectation  of $C_F/C_A   =   9/4$,  suggesting  that higher     order
corrections and non-perturbative effects are very important to understand the
measured value.  A  next-to-leading order  correction \cite{mueller}  in  MLLA
(Modified  Leading Log Approximation) at O$(\sqrt{\alpha_s})$ already
lowers the  prediction towards $r$ values  slightly below two and  exhibits a
small energy  dependence due to the  running of $\alpha_s$.  However  this is
still insufficient to explain the value of $r$ determined by the experiments.
Solutions    based   on    the   Monte    Carlo    method  give    a   better
approximation \cite{delphi_qg}.  The parton shower option  of the {\sc Jetset}
generator \cite{jetset} which uses   the Altarelli-Parisi splitting  functions
for     the   evolution  of  the   parton    shower  reduces  the theoretical
prediction \cite{delphi_qg} for $r$. At parton  level,  at 24 GeV jet  energy,
the expected value is $\sim$1.4 and it is further reduced to $\sim$1.3 if the
value of $r$ is computed after the fragmentation process. In both cases there
is a clear dependence of $r$ with the jet energy \cite{delphi_qg} which can be
parametrized using straight lines  with  slopes of  ${\Delta r}/{\Delta E}  =
(+90 \pm 3\mbox{(stat.)})\cdot 10^{-4} \ \mbox{GeV}^{-1}$ at parton level and
${\Delta   r}/{\Delta E}      =  (+76  \pm    2\mbox{(stat.)})\cdot  10^{-4}\
\mbox{GeV}^{-1}$ after fragmentation. The absolute  value of $r$ predicted at
parton level is however largely affected by the choice on the $Q_0$ parameter
(cut-off at which the parton evolution stops) but has negligible influence on
its relative  variation with the energy, i.e.,  the slope. The DELPHI analysis
uses symmetric and  non-symmetric  three-jet event configurations  with quark
and gluon jets  of variable energy, allowing  thus  all these  properties and
predictions to be tested. A value of $r=1.23\pm0.03\ \mbox{(stat.+syst.)}$ is
measured corresponding to an average  jet energy of  $\sim$27 GeV. The energy
dependence of $r$ is also suggested at 4$\sigma$ significance level, with a
fitted slope of ${\Delta r}/{\Delta E} = (+104\pm 25\mbox{(stat.+syst.)})\cdot
10^{-4}\ \mbox{GeV}^{-1}$.

{\small
\begin{figure}[hbt]
\begin{center}
\mbox{\epsfig{file=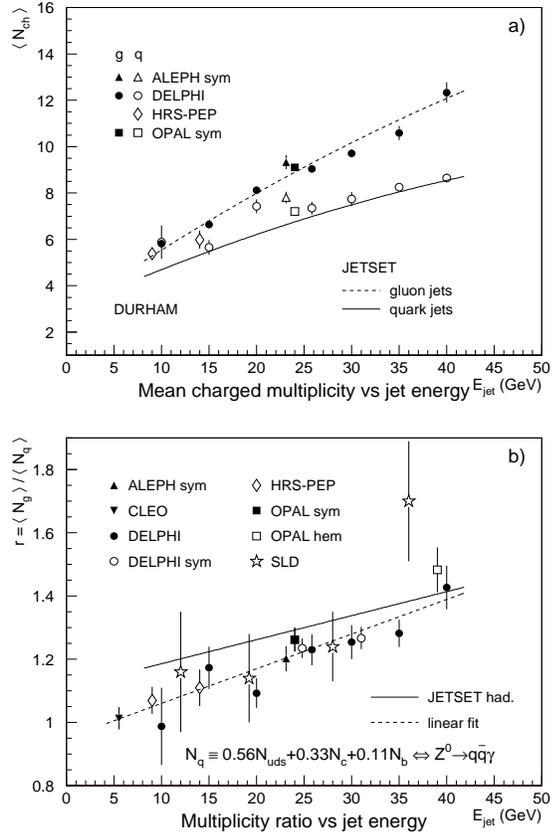,height=11.cm,width=7.5cm}}
\end{center}
%\caption{ (a) Mean charged multiplicity of quark and gluon jets as a function
%of the jet energy for a flavour composition of 33\% $c$'s and 11\% $b$'s. 
%(b) Multiplicity ratio $r$ as a function of the jet energy. Values corresponding 
%to the  same energy  are separated for having a better display.}
\caption{ (a) Mean charged multiplicity of quark and gluon jets and 
(b) multiplicity ratio $r$ as a function of the jet energy}
\label{fig:proc_mult}
\end{figure}
}

In a recent review \cite{bruseles} all published data from various 
experiments \cite{delphi_qg,opal_qg,aleph_qg,hrs_qg,sld_qg} were used  to 
perform a general study of $r$ as a function of the jet energy. At 
present, more data can be 
added to this comparison. These are the new analysed DELPHI data sample 
presented above and 
the most recent measurements of $r$ performed by CLEO \cite{cleo_qg} and 
OPAL \cite{opal_hem} at 4-7 GeV and 39 GeV average jet energies,
respectively.  The updated new DELPHI analysis incorporates two times 
more statistics than the previous analysis \cite{delphi_qg}, therefore, significantly
reduces the statistical errors. The analysis from CLEO compares the charged particle 
multiplicity in $\Upsilon(1S) \rightarrow gg\gamma$ decays
to that observed in $e^+e^- \rightarrow q\overline{q}\gamma$ just in the
continuum. This study does not rely on the Monte Carlo simulation to associate
the final hadrons to the initial partons and can consequently be fairly considered as
being model independent. The obtained value is 
$r=1.04\pm0.05\ \mbox{(stat.+syst.)}$. The OPAL analysis uses a new 
technique \cite{gary} which selects gluon jets at $\sim$39 GeV by 
dividing the events into two hemispheres. While one of these hemispheres
is required to contain two tagged quark jets, the other is left untouched 
being regarded as the gluon jet. The result from OPAL, expressed for only light 
$uds$-quarks, is $r_{uds}=1.55\pm0.07\ \mbox{(stat.+syst.)}$. As 
it can be observed in figure \ref{fig:proc_mult}.b  all these data agree  
with the predicted energy behaviour of \cite{delphi_qg,fodor,bruseles} when
the correction to the quark multiplicity to account for the same flavour 
composition is applied. In our case it is: 56\% $uds$'s, 33\% $c$'s 
and 11\% $b$'s. The OPAL number considering this quark mixture becomes  
$r=1.48\pm0.07\ \mbox{(stat.+syst.)}$.

All these results thus give evidence for an energy  dependence
of $r$. The measured increase is
\[
\frac{\Delta r}{\Delta E} = 
(+110 \pm 13\ \mbox{(stat.+syst.)})\cdot10^{-4}\  \mbox{GeV}^{-1},
\]
representing a $\sim$8$\sigma$ effect.

The measured value of $r$ remains  systematically lower than the {\sc Jetset}
prediction over the whole energy range, having an average value of
\[
r= 1.23\pm0.01\   (stat.)  \pm0.03\  (syst.),
\]
which corresponds to an average energy of $\sim$23 GeV. This ratio can be further
expressed as
\[
r_{uds}= 1.30 \pm0.01\ (stat.) \pm0.04\ (syst.), 
\]
if $r$ is computed only for the light $uds$-quarks, extracting the
$b$ and $c$ quark contribution  to  the quark  jet multiplicity.

The absolute value  of $r$ depends on the  reconstruction jet algorithm.  For
both  the {\sc Jade} and {\sc   Cone} schemes different  results are obtained
w.r.t.   the {\sc Durham} scheme \cite{delphi_qg,opal_qg}. This  is due to the
combined  effect   of   the   different  sensitivity  of   the    various jet
reconstruction  algorithms to  soft  particles  at  large angles   and of the
expected different  angular and energy spectra  of the emitted soft gluons in
the quark and gluon jets.  A precise deconvolution of  both effects is, at
present,  impossible \cite{ruso}.  This jet algorithm dependence of $r$ 
becomes however less apparent as the jet energy increases. The results from
OPAL \cite{opal_hem}, $r=1.48\pm0.07\ \mbox{(stat.+syst.)}$ and those from 
DELPHI \cite{delphi_qg} at $\sim$40 GeV presented in this conference,
$r=1.43\pm0.07\ \mbox{(stat.+syst.)}$ for {\sc Durham} and 
$r=1.52\pm0.11\ \mbox{(stat.+syst.)}$ for {\sc Jade}, agree within errors 
for the various methods and algorithms used. For the low energy 
interval, the {\sc Jade} and {\sc Durham} jet algorithms give a different 
description of the gluon jet properties \cite{delphi_qg}, although the 
{\sc Durham} algorithm is in better agreement to 
those, $model$ $independent$, results obtained by CLEO. Hence, the 
{\sc Durham} jet algorithm seems to be better suited to decribe the 
intermediate energy region than the {\sc Jade} algorithm is.

The interpretation of these results in combination with those obtained by 
OPAL \cite{opal_bg} and ALEPH \cite{aleph_bg} restrict the validity of the 
statement that gluon and $b$-quark jets have similar properties to the jet 
energy interval around 24 GeV and cannot be applied to the whole 
jet energy spectrum.  
  
\section{GLUON RADIATION IN $b$-QUARKS}

For many observable quantities at LEP energies, $\sqrt{s} \gessim$M$_Z$, 
quark mass effects usually appear in terms proportional to 
m$_q^2/$M$_Z^2$. This represents a $\sim$3\permil correction 
for m$_q$=m$_b$ which in most of the cases can be savely 
neglected. This argument, for instance is true for the total 
hadronic cross section \cite{val} but cannot be applied for the 
differential multi-jet cross sections that depend on 
the jet-resolution parameter, $y_c$. The reason being the
new scale, $E_c = M_Z \sqrt{y_c}$, introduced in the analysis
by the new variable which enhances the mass effects in the form
$m_b^2/E_c^2 = (m_b^2/M_Z^2)/y_c$. At $\sqrt{s} \approx$M$_Z$ 
the three-jet production rate for $b$-quarks is in fact suppressed 
by a factor $\sim5-10$\% w.r.t that of light 
quarks \cite{tor,val,delphi_ab}. This difference can then be 
expressed as a function of m$_b$ \cite{val} and, therefore, used to 
measure its value. 

\begin{figure}[ht]
\begin{center}
\mbox{\epsfig{file=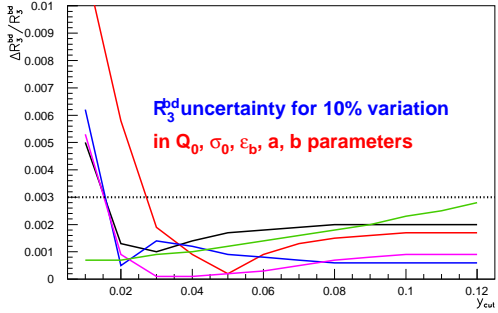,width=7.5cm} }
\end{center}
\caption{Relative systematics uncertainties in $R_3^{bd}$ due to fragmentation}
\label{fig:systematics}
\end{figure}

The experimental observation of such effects is however difficult 
and delicate because the effect is after all small and furthermore the  
correct theoretical framework to resolve the mass definition 
ambiguities is needed. This means that the observable has to be calculated
including mass effects at O($\alpha_s^2$). For this purpose a recent
calculation \cite{german} of the ratio of the normalized 
three-jet cross sections between $b$-quarks and light $uds$-quarks 

\[
R_3^{bd} \equiv \frac{\Gamma_{3j}^{Z^0\rightarrow b\bar{b}g}(y_c)/
                      \Gamma_{tot}^{Z^0\rightarrow b\bar{b}}}
                     {\Gamma_{3j}^{Z^0\rightarrow d\bar{d}g}(y_c)/
                      \Gamma_{tot}^{Z^0\rightarrow d\bar{d}}}  
\label{eq:r3bd}
\label{rtheta}
\]
has been performed

\begin{figure}[ht]
\begin{center}
\mbox{\epsfig{file=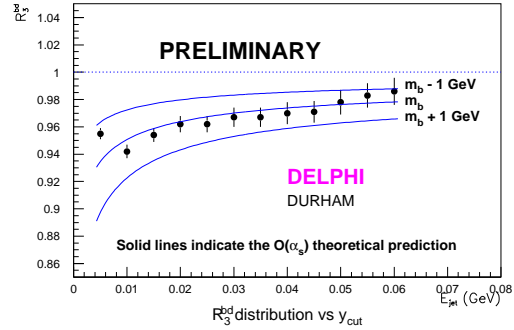,width=7.5cm} }
\end{center}
\caption{$R_3^{bd}$ distribution}
\label{fig:r3bd}
\end{figure}

The normalization in $R_3^{bd}$ to the total decay rates is introduced to 
cancel possible weak corrections depending on the top quark mass \cite{top}
and the ratio of the three-jet cross sections between $b$ and light 
$uds$-quarks minimizes uncertainties due to the hadronization process. In
figure \ref{fig:systematics} the dependence of these uncertainties w.r.t the
$y_{cut}$ is shown and seen not to exceed 3\permil for 
large enough values of $y_{cut}$. The $R_3^{bd}$ distribution corrected for detector and
fragmentation effects is also displayed in figure \ref{fig:r3bd}. The solid curves
drawn in the figure are the theoretical O($\alpha_s$) prediction in steps of 1
GeV. The values of m$_b$ used to produce these curves are meaningless since
they correspond to a calculation at  O($\alpha_s$). They can nevertheless be used to
evaluate the experimental precision assuming the difference between 
the theoretical curves remains similar to that at O($\alpha_s^2$). As can be 
observed the experimental error corresponds then to approximately 300 MeV for 
reasonably high values of $y_{cut}$.   
   
%%%%%%%%%%%%%%%%%%%%%%%%%%%%%%%

%\section*{References}

\end{document}